\documentclass[english,aps,preprint]{revtex4}
\usepackage[T1]{fontenc}
\usepackage[latin9]{inputenc}
\usepackage{amssymb}
\usepackage{esint}

\makeatletter
\@ifundefined{textcolor}{}
{%
 \definecolor{BLACK}{gray}{0}
 \definecolor{WHITE}{gray}{1}
 \definecolor{RED}{rgb}{1,0,0}
 \definecolor{GREEN}{rgb}{0,1,0}
 \definecolor{BLUE}{rgb}{0,0,1}
 \definecolor{CYAN}{cmyk}{1,0,0,0}
 \definecolor{MAGENTA}{cmyk}{0,1,0,0}
 \definecolor{YELLOW}{cmyk}{0,0,1,0}
}

\makeatother

\usepackage{babel}
\begin{document}

\title{Notes on Entropic Interpretation of Gravity}

\author{Qiao-Jun Cao}

\email{caoqiaojun@gmail.com}

\address{\emph{Department of physics, Shaoxing University, Shaoxing, 312000,
China}}
\begin{abstract}
Some essential conceptual aspects that will fill some logical gaps
of the frame to interpret the gravity as an entropic force was investigated,
we focus on some crucial issues that didn't emphasized in Verlinde's
original paper{[}arXiv:1001.0785{]}. This note explains the context
that holographic screen can be endowed with an entropy that proportional
to it's area and the meaning in using equipartition law in spacetime
thermodynamics, thermodynamic quantities such as entropy and temperature
are observer dependent is the crucial concept in explaining those
problems. Coarse graining will leave information in the gravitational
potential, which will connect different observer's point of views
for the same object. This will help us to understand the coarse graining
dependent definition of entropy and the nature of spacetime. It also
indicates the way that entropy bounds work, which is consistant with
Bekenstein entropy bound and holographic entropy bound. 
\end{abstract}
\maketitle

\section{Introduction}

Since the discovery of black hole thermodynamics by Bekenstein\cite{Bekenstein1973}
and Hawking\cite{Hawking1975}, the concept that gravity is an emergent
phenomenon has been widely approved. This idea that there should be
internal degrees of freedom exist for spacetime can be traced back
to the year 1968, when Sakharov found some similar behavior in the
elasticity and the spacetime dynamics\cite{sakharov2000vacuum}. The
similarity between gravitational force and Elastic force is not just
a coincidence, but have some profound connections. Elastic force is
not seem as a fundamental force, gravitational force should also not
fundamental, if elasticity and gravity originate from the same kind
of mechanism. Current gravity theory such as Einstein's general relativity
may be an emergent low-energy long-distance phenomenon that is insensitive
to the details of the underlying quantum theory of gravity. It will
be very exciting if we have an statistical description of the internal
degrees of freedom of spacetime, although this is just an extravagant
hope at present. 

In 2010, a paper by Erik P. Verlinde attracts tremendous attention.
He advocated the idea that gravity is an emergent phenomenon and asserted
that gravity is an entropic force in \cite{Verlinde:2010hp}. The
main idea of entropic interpretation of gravity is summarized as follows:
gravitional system have an microscopic structure, that is to say,
it is a thermodynamic system that all of the thermodynamic variables
can be endowed, gravity is just a statistical tendency to return to
a maximal entropy state. Based on some general assumptions, such as
the holographic princple, the equipartion law of energy and the Unruh
temperature formula, the Newton's law of gravity and the second law
of Newton can be derived.

The entropic interpretation of gravity give rise to a variety of debates,
the scientific community is divided into two sides, some people support
the idea, others don't, for the criticism, see \cite{Visser:2011jp}
and \cite{Wang:2012gc}. The entropic interpretation of gravity give
us a new perspective in the study of cosmology, the Friedmann equation
can be derived by using this frame\cite{Cai:2010hk,Shu:2010nv}. It
also have been used to study the dark energy\cite{Li:2010cj,Easson:2010av},
cosmological inflation and acceleration\cite{Cai:2010zw,Wang:2010jmb}.

In this note, we disscuss some essential conceptual aspects of this
paradigm that is still obscure in Verlinde's paper. As we have stated
in \cite{Cao:2014haa}, those aspects stated below should be further
investigated:
\begin{enumerate}
\item In Verlinde's paper \cite{Verlinde:2010hp}, the boundary of a gravitational
system is a suppositional holographic screen, the derivation of the
dynamic equation of gravity and the discussion of the emergence of
spacetime is all related to the holographic screen. However, there
is no explicit explanation in \cite{Verlinde:2010hp} that how can
we endow the holographic screen a temperature and an entropy. It should
be noticed that the temperature and entropy of a gravitational system
is observer dependent, since we can only endow a temperature and an
entropy to an observer dependent horizon, when it comes to a holographic
screen, we need an explanation to fullfills the logical gap.
\item The equipartition law is a key assumption in deriving Newton's law
and Einstein's field equation in \cite{Verlinde:2010hp}, but we still
don't kown the implication of being used to spacetime thermodynamics
and the context of its application. 
\item In \cite{Verlinde:2010hp}, holographic screen is viewed as the boundary
that separate the spacetime into an emergent part and a non-existent
part. It is of no sense to separate the space into two regions with
a suppositional screen, and said that one part is existed and the
other is not. What is the real edge that separate the emergent spacetime
region from which has not emerged yet?
\end{enumerate}
Those problems stated above are crucial, we want to clarify those
problems in this note. It is surprised that they are all close related
to the properties of horizon, and thermodynamic quantities are observer
dependent in the thermodynamic description of spacetime. All those
problems will be automatically resolved after we made some conceptual
change about the spacetime. 

Some authors have developed a formal thermodynamic first law on holographic
screens with spherical symmetry\cite{chen2010first,wu2010thermodynamics},
but the physical implication of their results is ambiguous. We found
that out result will provide an explanation to what is the explicit
physical meaning of the thermodynamic parameters used in their formula. 

This note will be organized as follow: In sec.\ref{sec:horizon and screen},
the important role of horizon played in spacetime thermodynamics was
discussed, and we explain the context that holographic screen can
be endowed with an entropy that proportional to it's area. In sec.\ref{sec:The-observer-dependence},
we illustrate the meaning and conditions when we apply the thermodynamic
equipartition law to spacetime thermodynamics. In sec.\ref{sec:Coarse-graining-of},
we discuss coarse graining in gravitational system and it's implication
in understand the coarse graining dependent definition of entropy
and the entropy bounds. Discussions and conclusions were made in sec.\ref{sec:Discussions-and-conclusions}.

\section{Horizons and Holographic Screens\label{sec:horizon and screen}}

In general relativity, light cones are effected by gravity, and it
follows that there will exist observers who do not have access to
part of the spacetime, that is to say those observers will perceive
horizons. It should be noticed that all horizons are observer dependent,
and in general relativity all observers are set on an equal footing
because this is the crucial reason for Einstein to establish general
relativity, we should also treat all horizons equally in the study
of spacetime thermodynamics\cite{Padmanabhan:2003gd,Padmanabhan:2009ey}. 

Compared to ordinary screens in spacetime, horizons have distinctive
features. It has a key property that it can block information from
the corresponding observer which ordinary screens don't have, it is
this property of horizon that make us to endow an entropy to the horizon
for the observers who perceive the horizon. Supposing energy flows
across the horizon, the entropy in the region accessible to the observer
can decrease because the entropy carried by the energy flow is not
accessible to the observer any longer, if the horizon doesn't have
an entropy, the second law of thermodynamics will be violated. Therefore,
horizon should have an entropy and a corresponding temperature, and
the field equation of gravity can be derived by demand the Clausius
relation is hold on horizons\cite{Padmanabhan:2009ey,jacobson1995thermodynamics}.

It seems unnatural to endow an entropy to ordinary screens, because
it is not necessary to require ordinary screens have an entropy to
avoid the violation of the second law of thermodynamics, the information
carried by energy flow across a ordinary screen is still approachable
for the corresponding observer. In fact, the distinction between horizons
and ordinary screens is artificial, the black hole horizon is just
an ordinary screen from the point of view of an free falling observer,
and we can construct local Rindler horizons to cover every small patch
of any ordinary screens\cite{padmanabhan2004entropy}, is it a horizon
or not is all depend on which kinds of observers watch it. 

It should be noticed that observer dependent is a crucial concept
in the discussion of following paragraphs. The holographic screen
is introduced by susskind to illustrate the holographic principle\cite{Susskind:1994vu}.
For an observer a holographic screen could be a horizon or an ordinary
screen that encompass the horizon\cite{Bousso:1999xy,Bousso:1999cb}.
The holographic principle\cite{'tHooft:1993gx,Susskind:1994vu} states
that a macroscopic region of space and everything inside it can be
represented by a boundary theory living on the boundary of the region.
The strongest supporting evidence for the holographic principle comes
from black hole physics and the AdS/CFT correspondence, in those two
cases the number of degrees of freedom of the boundary surface agrees
with the number of physical degrees of freedom contained in the bulk.
The entropy of a Schwarzschild black hole, $S_{BH}=A_{horizon}/4$,
precisely saturates the holographic entropy bound. In this sense,
if we view the black hole as an isolate system, a black hole is the
most entropic object one can put inside a given spherical surface.
This is not surprising since the gravitaional evolution can be viewed
as a thermodynamic process for a system to reach an equilibrium state
that the holographic principle attained\cite{Cao:2014haa}. For complete
weakly self-gravitating physical system surrounded by a spacelike
surface with area $A$, entropy of the system, $S<A/4$, will not
saturate the bound\cite{Susskind:1994vu}, notice that this surface
is not a horizon from the point of view of the observers that view
the system as an ordinary one. We may take the surface as a holographic
screen, and there possibly exits a theory on it --though hard to establish,
and may have peculiar properties, don't like the AdS/CFT-- that dual
to the bulk theory for the system. The difficulty in establishing
such theory probably have something to do with redundant degrees of
freedom on holographic screen, we will discuss it in sec.\ref{sec:Coarse-graining-of}. 

For a gravitating system enclosed by a holographic screen, we can
construct local Rindler horizons for every observers placed on the
holographic screen who will experience an acceleration $a$ produced
by the gravitational body, then, we can attribute an Unruh temperature
$T=\frac{\hbar a}{2\pi k_{B}c}$ to it, and attribute an entropy 
\begin{equation}
S_{screen}=\frac{c^{3}}{4G\hbar}\int_{\mathcal{S}}dA\label{eq:holographic entropy}
\end{equation}
 to the holographic screen. Obviously, this amount of entropy will
violate the Bekenstein entropy bound\cite{bekenstein1981universal,Verlinde:2010hp},
we will clarify this problem in sec.\ref{sec:The-observer-dependence}
and \ref{sec:Coarse-graining-of}. One can think about the holographic
screen as a storage device for information, as stated in\cite{Verlinde:2010hp},
we call the fundamental degree of freedom (or fundamental atom) on
holographic screen bit, note that the bit here is not the unit used
to measure information. If we assume the holographic principle holds,
the total number of bits $N$ on holographic screen is proportional
to it's area $A$, that is 
\begin{equation}
N=\frac{Ac^{3}}{G\hbar}.\label{eq:total bits}
\end{equation}
The degrees of freedom is consistent with entropy formula eq.\ref{eq:holographic entropy},
and the dynamics of those bits on holographic screen is governed by
the unknown dual theory mentioned above, which also rules how to store
information by bits. Note that, for an observer, the holographic principle
doesn't set the horizon and ordinary screen on an equal footing. The
maximal storage capability of a holographic screen equals the total
number of bits $N$ when the holographic screen is a horizon, that
is encode one bit information by one fundamental degree of freedom,
which happens when the holographic entropy bound is saturated. When
the holographic entropy bound is not saturated and the holographic
screen is not a horizon, more than one bits are used to encode one
bit information, which is stored and processed in an unknown coarse
graining way.

\section{The observer dependence of equipartition law\label{sec:The-observer-dependence}}

Horizon can block information from the corresponding observer and
separate the spacetime into two different parts, one part is accessible
and the other is not. In this sense, the horizon plays as a boundary
of a gravitaional system, the system is not separated from the other
by a suppositional screen, but by a causality barrier, and the ``system''
contain the degrees of freedom beyond the horizon which can have a
dual description on the horizon. 

A complete understand of the fundamental degree of freedom requires
a consistent theory of quantum gravity, which has so far proved elusive.
However, just as semiclassical analysis such as the Bohr model was
important in the early development of quantum mechanics, a similar
approach may be helpful in understanding some of the microscopic features
of spacetime. To gain some intuitive understanding of the spacetime
atom, let us do some semiclassical analysis for a Schwarzschild black
hole with Schwarzschild radius $R$ and mass $M$. The number of the
microscopic degrees of freedom of the black hole is $N=A_{horizon}$.
If we divide the black hole mass $M$ evenly over the microscopic
degrees of freedom, each spacetime atom will have an energy $E_{\gamma}=\frac{1}{8\pi R}$,
and the Compton wavelength of it is $\lambda_{\gamma}\sim R$. This
means the black hole horizon behaves like a ``box'' that confi{}ne
the whole spacetime atom within it. Modes with Compton wavelength
$\gtrsim R$ can not be confined in the black hole, and do not contribute
to the entropy of the black hole; while modes with Compton wavelength
$\lesssim R$ is not the most effective way to increase the entropy
of the black hole, this is conflict with that black hole is a most
entropic object. We speculate that the formation of a black hole is
to distribute ordinary form of matter in a manipulative way with efficiency
to make it as a most entropic object from the point of view of a proper
observer. There is a gap in magnitude between black hole entropy and
ordinary entropy, it was proven in \cite{chen2008entropy} that entropy
bound for ordinary system is bounded by $A^{\frac{3}{4}}$ , it should
undergo a peculiar process when the black hole is formed, we will
come back to this issue in sec.\ref{sec:Coarse-graining-of}.

The thermodynamic description of spacetime is independent of the exact
nature of the degrees of freedom, although we don't have definite
knowledge about atoms of spacetime, we can apply the thermodynamic
laws to gravitational system. In Verlinde's original paper \cite{Verlinde:2010hp},
the equipartition law play a crucial role in deriving Newtion's law
and Einstein's field equation. This is analogous to what we did to
a Schwarzschild black hole, the total energy $E$ for a system is
divided evenly over the microscopic degrees of freedom of spacetime
$N$:
\begin{equation}
E=\frac{1}{2}Nk_{B}T\label{eq:equipartition law}
\end{equation}
where $k_{B}$ is Boltzman's constant and $T$ represents the temperature
of the system. We can get this result if we attribute an energy $(1/2)k_{B}T$
to each microscopic degree of freedom of spacetime. 

In Verlinde's paper \cite{Verlinde:2010hp}, the meaning of the temperature
$T$ is obscure. When $T$ equals the Unruh temperature $T=\frac{\hbar a}{2\pi k_{B}c}$,
we can get the second law of Newton: $F=ma$; if $T$ is explained
as the temperature of the holographic screen and also equals the Unruh
temperature, the following entropy formula (\ref{eq:holographic entropy})
of the screen will violate the Bekenstein entropy bound as we stated
in eq.(\ref{eq:holographic entropy}). What are the reasons of this
discrepancy ? In fact, this discrepacncy could be avoid after some
issues of this problem are clarified. The entropy of a complete weakly
self-gravitating physical system is bounded by the Bekenstein entropy
bound\cite{bekenstein404042does}, and the Bekenstein entropy bound
has been explicitly shown to hold in wide classes of equilibrium systems\cite{schiffer1989proof},
the system considered in Verlinde's paper is unlikely to violate the
Bekenstein entropy bound. On the other hand, it is absurd to conclude
that the temperature of the holographic screen is not the Unruh temperature,
although the Unruh temperature formula is not necessary in deriving
Newton's law of gravitation $F=G\frac{Mm}{R^{2}}$, if this is true,
the gravitational force should not an ordinary force which obey the
second law of Newton, since the origin of the temperature are different. 

We should note that the temperature and entropy are observer dependent
quantities in the thermodynamical description of gravity\cite{padmanabhan2010thermodynamical}.
An observer falling into a black hole will ascribe different thermodynamic
properties to the black hole compared to an observer who is remaining
stationary outside the horizon. It is also true that an inertial observer
will attribute different temperature and entropy to the Minkowski
vacuum compared to a Rindler observer. The entropy of the holographic
screen $S_{screen}=\frac{c^{3}}{4G\hbar}\int_{\mathcal{S}}dA$ is
associate with the observers who will experience an Unruh temperature
$T=\frac{\hbar a}{2\pi k_{B}c}$, those observers will ascribe this
entropy to the holographic screen because they experience an acceleration
produced by the gravitational body and we can construct local Rindler
horizon for any small patch of the holographic screen, any other observers
will not attribute the same amount of entropy to the holographic screen.

When one does quantum field theory in curved spacetime, ``particle''
become an observer dependent notion, it is not surprise to associate
different amount of entropy with a gravitational system for different
kinds of observers. In \cite{marolf2004notes}, the authors showed
that the entropy associated with a ordinary localized object in flat
and otherwise empty space is not an invariant quantity defined by
the system alone, but rather depends on which observer we ask to measure
it. From the inertial observer's point of view, the entropy of an
object with $n$ possible microstates and energy $\delta E$ is $\delta S_{inertial}=\ln n$,
form the point of view of the Rindler observer, the entropy of the
object with the same resolution is
\begin{equation}
\delta S_{Rindler}=\frac{\delta E}{T},\label{eq:first law}
\end{equation}
 we see the Rindler entropy is not necessary equals to the inertial
entropy, the relation between those two kinds of entropy is still
in the dark, but we have reasonable ground to believe this problem
will be uncovered after we understand the nature of spacetime, in
section \ref{sec:Coarse-graining-of}, we will argue that it has something
to do with coarse gaining dependent of entropy. From the Rindler observer's
point of view the relation (\ref{eq:first law}) can be interpreted
as the first law of thermodynamics on holographic screens in some
sense\cite{chen2010first}, if an object carries energy $\delta E$
falling into the local Rindler horizon, the associated change of the
holographic screen entropy will be $\frac{\delta E}{T}$.

Now, the obscureness in apply the equipartition law to gravitational
system discussed above can be cleared up. When we want to measure
the length of a object, we need a ruler,which means we have to use
certain quantity of length as the standard length, similarly, we can
apply the equipartition law to a gravitational system because spacetime
have microscopic degrees of freedom. We have to conclude that the
application for the equipartition law is also observer dependent,
since the fundamental energy $k_{B}T/2$ is temperature dependent,
and different $T$ correspond to different Rindler observer. We should
note that in Schwarzschild spacetime a infinity observer is immersed
in a Hawking radiation with temperature $T=\frac{1}{8\pi M}=\frac{1}{4\pi R}$,
this observer will have a ``ruler'' with standard fundamental energy
$\frac{1}{2}k_{B}T=\frac{k_{B}}{8\pi R}$, it happens that each fundamental
degree of freedom have the same energy as we discussed above. In the
above case used in Verlinde's paper, Rindler observers in different
spherical surfaces will experience different Unruh temperatures, and
will have different ``ruler'', so they will divide the same system
into different amount of degrees of freedom. Eq.(\ref{eq:equipartition law})
can be thought as the integrated form of eq.(\ref{eq:first law})\cite{chen2010first}.

\section{Coarse graining of entropy and Entropy bound\label{sec:Coarse-graining-of}}

We want to point out that we can't talk about the entropy of a holographic
screen without regarding to the circumstance of physical system. For
a inertial observer, it is ridiculous to endow an entropy to a screen
in Minkowski spacetime that is equal to it's area, similarly, for
a observer placed at infinity, the entropy of a Schwarzschild black
hole will not change when we go away from the horizon and choose a
holographic screen that contains the horizon inside it, the observer
dependent entropy is always equals to $\frac{A_{horizon}}{4}$. We
therefore conclude that information can be stored on screens, and
the amount of information that stored on it is determined by circumstance
of the whole system and the corresponding observer that measures it,
rather than the screen itself. Generally speaking, the capacity of
the screen to store information is no less than information stored
on the whole system surrounded by it unless the holographic screen
happened to be the horizon of the corresponding observer.

Entropy is an extremely subtle concept in general relativity, a proper
framework for general discussion of entropy is still lack\cite{wald1999gravitation}.
We have known that entropy is an observer dependent quantity and it's
definition is coarse graining dependent, a question follows, what
is the relationship between observers and coarse graining? We will
see that the entropic force interpretation of gravity will give us
some clue about this problem.

Verlinde's entropic force interpretation of gravitational force is
really of some attractive properties and profound meanings. It is
the first time that uncovered the origin of the gravitational force
and inertia, and the mechanism of gravity is really clear and imaginable.
Lots of interesting application of Verlinde's proposal have been studied,
in \cite{Li:2010cj}, the authors showed that the UV/IR relation proposed
by Cohen et al., as well as holographic dark energy can be derived
from the entropic force formalism. The crucial equation used in \cite{Li:2010cj}
is a relation between entropy $S$, used bits $N$ and Newton potential
$\Phi$:
\begin{equation}
\frac{S}{N}=-k_{B}\frac{\Phi}{2c^{2}}.\label{eq:coarse graining}
\end{equation}
The Newton potential $\Phi$ can be identified as a coarse graining
variable, and $0\leqslant-\frac{\Phi}{2c^{2}}$$\leqslant\frac{1}{4}$.
This relation is extremely consistent with our illustration in section
\ref{sec:horizon and screen}, because it makes it clear that the
number of bits on the holographic screen which are used to dually
describe the object in the bulk can be either equal to or larger than
the entropy of the bulk object. Consider two observers placed on two
different equipotential holographic screens with Newton potential
$\Phi_{1}$ and $\Phi_{2}$, they will experience different acceleration
produced by the bulk object, and will endow two observer dependent
entropy $S_{1}$ and $S_{2}$ to the gravitating object, according
to eq.(\ref{eq:coarse graining}), $S_{1}$ and $S_{2}$ satisfy 
\begin{equation}
\frac{S_{1}}{S_{2}}=\frac{\Phi_{2}}{\Phi_{1}},
\end{equation}
we see, this relation indicate a relationship between observers and
coarse graining, in other words, Newton potential $\Phi$ is a phenomenological
parameter, which keeps track of the message for a coarse graining
description of the bulk object on different holographic screens. Observers
at rest on different holographic screens will experience different
acceleration produced by the bulk object, different observers will
endow different entropy to the bulk object, as we discussed in section
\ref{sec:The-observer-dependence},and those different entropy correspond
to different observers are link by gravitational potential. 

We can use another example to illustrate gravitational potential measures
the amount of coarse graining. Consider a object with proper energy
$E$ that can surrounded by a sphere with radius $r$ in a Schwarzschild
black hole background, with Schwarzschild radius $R$ and mass $M$.
When place the object at infinity, a infinity observer is immersed
in a Hawking radiation with temperature $T=\frac{1}{8\pi M}=\frac{1}{4\pi R}$,
for this observer each microscopic degree of freedom will attribute
an energy $\frac{1}{2}k_{B}T=\frac{k_{B}}{8\pi R}$, note that a fundamental
degree of freedom of the black hole will have the same energy if we
divide $M$ evenly over $N_{horizon}=A$, then, the total number microscopic
degrees of freedom of the object at infinity is $N_{infinity}=\frac{8\pi ER}{k_{B}}$,
and the entropy is $S_{infinity}=2\pi ER$. On the other hand, when
we place the object just hanging out the black hole, that is the proper
distance from the center of the object to the horizon is $r$, the
infinity observer will attribute an energy $\frac{Er}{2R}=\frac{Er}{4M}$
to the object, where $V=\frac{r}{2R}$ is the red shift factor, when
it captured by the black hole, the entropy change of the black hole
is $\delta S_{BH}=8\pi M\delta M=2\pi Er$, then, the entropy of the
object should be bounded by $\delta S_{BH}$, that is $S_{hanging}\leq2\pi Er$.
We found that for an infinity observer in Schwarzschild black hole
background, the same object placed in different location that have
different gravitational potential will be attributed different entropy.
This means when an object is in a background of gravity, the object
is not a isolate system, it's entropy is determined by the observer
as well as the gravitational background. 

Next, we want to show that the eq.(\ref{eq:coarse graining}) is consistent
with the entropy bound proposed both by Bekenstein\cite{bekenstein1981universal}
and Susskind\cite{'tHooft:1993gx}. First, let's go back to the problem
noticed by Verlinde in Sec. 6.4 of Ref.\cite{Verlinde:2010hp}, which
state that if one endow an Unruh temperature to a holographic screen
that is not a horizon, the following entropy formula $S_{screen}=\frac{c^{3}}{4G\hbar}\int_{\mathcal{S}}dA$,
will violate the Bekenstein entropy bound. This problem can be clarified,
as we state above different observers will attribute different amount
of entropy to the same object, from the Rindler observer's point of
view the object is no longer an isolate system in ordinary asymptotically
flat spacetime, it is not a contradiction that a inertial observer
who looks the object as a complete, weakly self-gravitating, isolate
system in ordinary asymptotically flat spacetime will endow it a different
amount of entropy. 

Now, let's explain the consistency of entropic interpretation of gravity
and entropy bound more explicit. Using eq.(\ref{eq:coarse graining})and(\ref{eq:total bits}),
we get
\begin{equation}
S=-k_{B}\frac{\Phi}{2c^{2}}\frac{Ac^{3}}{G\hbar}\leqslant\frac{Ak_{B}c^{3}}{4G\hbar},\label{eq:susskind bound}
\end{equation}
we use the fact that the maximum value of the ratio $-\Phi/2c^{2}$
is $1/4$ when the maximum coarse graining happens at horizons. This
is just Susskind's holographic entropy bound, and we should also note
that this is from the point of view of the observers that looks the
object as a complete, weakly self-gravitating, isolate system in ordinary
asymptotically flat spacetime.

If we use the equipartition rule $E=\frac{1}{2}Nk_{B}T$, then, eq.(\ref{eq:coarse graining})
can be write as:
\begin{equation}
S=-k_{B}\frac{\Phi}{2c^{2}}\frac{2E}{k_{b}T}.
\end{equation}
 Due to the Unruh effect, for the system with mass $M=E/c^{2}$ which
can be surrounded by a sphere of radius $R$, it is argue that $T\geqslant\frac{1}{8\pi M}\geqslant\frac{1}{4\pi R}$\cite{Abreu:2010sc}.
The Bekenstein entropy bound $S\leqslant2\pi ER$ is followed straightforward.

\section{Discussions and conclusions\label{sec:Discussions-and-conclusions}}

The entropic interpretation of gravity present in Verlinde's paper
is a very provocative idea, this is a good start, but the theory is
also very incomplete, first of all, these idea should be recast in
more precise way, and we still have a long journey to go to build
a complete gravity theory. 

In this small note, we present some explanations to some essential
conceptual aspects that will fill some logical gap in interpreting
the gravity as an entropic force which is missed in Verlinde's origin
paper, and we think this is a small step to get a more precise theory.
We have argued that thermodynamic quantities are observer dependent
is very important concept in studying spacetime thermodynamic. In
curved spacetime, energy is a observer dependent quantity, and particles
also become an observer dependent notion, observers can disagree on
the entropy and temperature of a system. Horizon is observer dependent
and the thermodynamic laws established in spacetime thermodynamic
are crucial related to it, because the entropy and temperature can
not be defined without horizon, moreover, horizon plays as a edge
of a gravitational thermodynamic system. The entropic interpretation
of gravity give us some clue to understand that coarse graining will
leave informations in the gravitational potential, this message can
help us to link different observer's point of views about the same
object. 

From the point view of Verlinde, gravitational force is caused by
entropy gradients when locations of material bodies changes, which
means, gravity is a statistical tendency to return to a maximal entropy
state. We know every object carries a amount of entropy, so the process
in tending to the maximal entropy state is a process to rearrange
the information, this is closely related the maximum speed of process
information for a system. We will investigate the relation between
the entropic interpretation of gravity and Margolus-Levitin Theorem
in future.

\begin{acknowledgments}
This work was supported by the NNSF of China under Grant No.11247306
and the NSF of Zhejiang Province of China under Grant No.LQ13A050002.
I would like to thank Cheng-Zhou Liu and Xuejun Yang for helpful comments
and suggestions.\bibliographystyle{unsrt}
\bibliography{entropic}

\begin{thebibliography}{31}
\expandafter\ifx\csname natexlab\endcsname\relax\def\natexlab#1{#1}\fi
\expandafter\ifx\csname bibnamefont\endcsname\relax
  \def\bibnamefont#1{#1}\fi
\expandafter\ifx\csname bibfnamefont\endcsname\relax
  \def\bibfnamefont#1{#1}\fi
\expandafter\ifx\csname citenamefont\endcsname\relax
  \def\citenamefont#1{#1}\fi
\expandafter\ifx\csname url\endcsname\relax
  \def\url#1{\texttt{#1}}\fi
\expandafter\ifx\csname urlprefix\endcsname\relax\def\urlprefix{URL }\fi
\providecommand{\bibinfo}[2]{#2}
\providecommand{\eprint}[2][]{\url{#2}}

\bibitem[{\citenamefont{Bekenstein}(1973)}]{Bekenstein1973}
\bibinfo{author}{\bibfnamefont{J.~D.} \bibnamefont{Bekenstein}},
  \bibinfo{journal}{Phys. Rev.} \textbf{\bibinfo{volume}{D7}},
  \bibinfo{pages}{2333} (\bibinfo{year}{1973}).

\bibitem[{\citenamefont{Hawking}(1975)}]{Hawking1975}
\bibinfo{author}{\bibfnamefont{S.~W.} \bibnamefont{Hawking}},
  \bibinfo{journal}{Commun. Math. Phys.} \textbf{\bibinfo{volume}{43}},
  \bibinfo{pages}{199} (\bibinfo{year}{1975}).

\bibitem[{\citenamefont{Sakharov}(2000)}]{sakharov2000vacuum}
\bibinfo{author}{\bibfnamefont{A.}~\bibnamefont{Sakharov}},
  \bibinfo{journal}{General Relativity and Gravitation}
  \textbf{\bibinfo{volume}{32}}, \bibinfo{pages}{365} (\bibinfo{year}{2000}).

\bibitem[{\citenamefont{Verlinde}(2010)}]{Verlinde:2010hp}
\bibinfo{author}{\bibfnamefont{E.~P.} \bibnamefont{Verlinde}}
  (\bibinfo{year}{2010}), \eprint{1001.0785}.

\bibitem[{\citenamefont{Visser}(2011)}]{Visser:2011jp}
\bibinfo{author}{\bibfnamefont{M.}~\bibnamefont{Visser}},
  \bibinfo{journal}{JHEP} \textbf{\bibinfo{volume}{1110}}, \bibinfo{pages}{140}
  (\bibinfo{year}{2011}), \eprint{1108.5240}.

\bibitem[{\citenamefont{Wang}(2012)}]{Wang:2012gc}
\bibinfo{author}{\bibfnamefont{T.}~\bibnamefont{Wang}} (\bibinfo{year}{2012}),
  \eprint{1211.5722}.

\bibitem[{\citenamefont{Cai et~al.}(2010{\natexlab{a}})\citenamefont{Cai, Cao,
  and Ohta}}]{Cai:2010hk}
\bibinfo{author}{\bibfnamefont{R.-G.} \bibnamefont{Cai}},
  \bibinfo{author}{\bibfnamefont{L.-M.} \bibnamefont{Cao}}, \bibnamefont{and}
  \bibinfo{author}{\bibfnamefont{N.}~\bibnamefont{Ohta}},
  \bibinfo{journal}{Phys.Rev.} \textbf{\bibinfo{volume}{D81}},
  \bibinfo{pages}{061501} (\bibinfo{year}{2010}{\natexlab{a}}),
  \eprint{1001.3470}.

\bibitem[{\citenamefont{Shu and Gong}(2011)}]{Shu:2010nv}
\bibinfo{author}{\bibfnamefont{F.-W.} \bibnamefont{Shu}} \bibnamefont{and}
  \bibinfo{author}{\bibfnamefont{Y.}~\bibnamefont{Gong}},
  \bibinfo{journal}{Int.J.Mod.Phys.} \textbf{\bibinfo{volume}{D20}},
  \bibinfo{pages}{553} (\bibinfo{year}{2011}), \eprint{1001.3237}.

\bibitem[{\citenamefont{Li and Wang}(2010)}]{Li:2010cj}
\bibinfo{author}{\bibfnamefont{M.}~\bibnamefont{Li}} \bibnamefont{and}
  \bibinfo{author}{\bibfnamefont{Y.}~\bibnamefont{Wang}},
  \bibinfo{journal}{Phys. Lett.} \textbf{\bibinfo{volume}{B687}},
  \bibinfo{pages}{243} (\bibinfo{year}{2010}), \eprint{1001.4466}.

\bibitem[{\citenamefont{Easson et~al.}(2011)\citenamefont{Easson, Frampton, and
  Smoot}}]{Easson:2010av}
\bibinfo{author}{\bibfnamefont{D.~A.} \bibnamefont{Easson}},
  \bibinfo{author}{\bibfnamefont{P.~H.} \bibnamefont{Frampton}},
  \bibnamefont{and} \bibinfo{author}{\bibfnamefont{G.~F.} \bibnamefont{Smoot}},
  \bibinfo{journal}{Phys.Lett.} \textbf{\bibinfo{volume}{B696}},
  \bibinfo{pages}{273} (\bibinfo{year}{2011}), \eprint{1002.4278}.

\bibitem[{\citenamefont{Cai et~al.}(2010{\natexlab{b}})\citenamefont{Cai, Liu,
  and Li}}]{Cai:2010zw}
\bibinfo{author}{\bibfnamefont{Y.-F.} \bibnamefont{Cai}},
  \bibinfo{author}{\bibfnamefont{J.}~\bibnamefont{Liu}}, \bibnamefont{and}
  \bibinfo{author}{\bibfnamefont{H.}~\bibnamefont{Li}},
  \bibinfo{journal}{Phys.Lett.} \textbf{\bibinfo{volume}{B690}},
  \bibinfo{pages}{213} (\bibinfo{year}{2010}{\natexlab{b}}),
  \eprint{1003.4526}.

\bibitem[{\citenamefont{Wang}(2010)}]{Wang:2010jmb}
\bibinfo{author}{\bibfnamefont{Y.}~\bibnamefont{Wang}} (\bibinfo{year}{2010}),
  \eprint{1001.4786}.

\bibitem[{\citenamefont{Cao}(2014)}]{Cao:2014haa}
\bibinfo{author}{\bibfnamefont{Q.-J.} \bibnamefont{Cao}}
  (\bibinfo{year}{2014}), \eprint{1401.1452}.

\bibitem[{\citenamefont{Chen and Li}(2010)}]{chen2010first}
\bibinfo{author}{\bibfnamefont{Y.}~\bibnamefont{Chen}} \bibnamefont{and}
  \bibinfo{author}{\bibfnamefont{J.}~\bibnamefont{Li}}, \bibinfo{journal}{Arxiv
  preprint arXiv:1006.1442}  (\bibinfo{year}{2010}).

\bibitem[{\citenamefont{Wu et~al.}(2010)\citenamefont{Wu, Ge, Zhang, and
  Yang}}]{wu2010thermodynamics}
\bibinfo{author}{\bibfnamefont{S.}~\bibnamefont{Wu}},
  \bibinfo{author}{\bibfnamefont{X.}~\bibnamefont{Ge}},
  \bibinfo{author}{\bibfnamefont{P.}~\bibnamefont{Zhang}}, \bibnamefont{and}
  \bibinfo{author}{\bibfnamefont{G.}~\bibnamefont{Yang}},
  \bibinfo{journal}{Arxiv preprint arXiv:1008.2072}  (\bibinfo{year}{2010}).

\bibitem[{\citenamefont{Padmanabhan}(2005)}]{Padmanabhan:2003gd}
\bibinfo{author}{\bibfnamefont{T.}~\bibnamefont{Padmanabhan}},
  \bibinfo{journal}{Phys. Rept.} \textbf{\bibinfo{volume}{406}},
  \bibinfo{pages}{49} (\bibinfo{year}{2005}), \eprint{gr-qc/0311036}.

\bibitem[{\citenamefont{Padmanabhan}(2009)}]{Padmanabhan:2009ey}
\bibinfo{author}{\bibfnamefont{T.}~\bibnamefont{Padmanabhan}}
  (\bibinfo{year}{2009}), \eprint{0910.0839}.

\bibitem[{\citenamefont{Jacobson}(1995)}]{jacobson1995thermodynamics}
\bibinfo{author}{\bibfnamefont{T.}~\bibnamefont{Jacobson}},
  \bibinfo{journal}{Physical Review Letters} \textbf{\bibinfo{volume}{75}},
  \bibinfo{pages}{1260} (\bibinfo{year}{1995}),
  \urlprefix\url{arXiv:hep-th/9504004}.

\bibitem[{\citenamefont{Padmanabhan}(2004)}]{padmanabhan2004entropy}
\bibinfo{author}{\bibfnamefont{T.}~\bibnamefont{Padmanabhan}},
  \bibinfo{journal}{Classical and Quantum Gravity}
  \textbf{\bibinfo{volume}{21}}, \bibinfo{pages}{4485} (\bibinfo{year}{2004}).

\bibitem[{\citenamefont{Susskind}(1995)}]{Susskind:1994vu}
\bibinfo{author}{\bibfnamefont{L.}~\bibnamefont{Susskind}},
  \bibinfo{journal}{J. Math. Phys.} \textbf{\bibinfo{volume}{36}},
  \bibinfo{pages}{6377} (\bibinfo{year}{1995}), \eprint{hep-th/9409089}.

\bibitem[{\citenamefont{Bousso}(1999{\natexlab{a}})}]{Bousso:1999xy}
\bibinfo{author}{\bibfnamefont{R.}~\bibnamefont{Bousso}},
  \bibinfo{journal}{JHEP} \textbf{\bibinfo{volume}{07}}, \bibinfo{pages}{004}
  (\bibinfo{year}{1999}{\natexlab{a}}), \eprint{hep-th/9905177}.

\bibitem[{\citenamefont{Bousso}(1999{\natexlab{b}})}]{Bousso:1999cb}
\bibinfo{author}{\bibfnamefont{R.}~\bibnamefont{Bousso}},
  \bibinfo{journal}{JHEP} \textbf{\bibinfo{volume}{06}}, \bibinfo{pages}{028}
  (\bibinfo{year}{1999}{\natexlab{b}}), \eprint{hep-th/9906022}.

\bibitem[{\citenamefont{'t~Hooft}(1993)}]{'tHooft:1993gx}
\bibinfo{author}{\bibfnamefont{G.}~\bibnamefont{'t~Hooft}}
  (\bibinfo{year}{1993}), \eprint{gr-qc/9310026}.

\bibitem[{\citenamefont{Bekenstein}(1981)}]{bekenstein1981universal}
\bibinfo{author}{\bibfnamefont{J.}~\bibnamefont{Bekenstein}},
  \bibinfo{journal}{Physical Review D} \textbf{\bibinfo{volume}{23}},
  \bibinfo{pages}{287} (\bibinfo{year}{1981}).

\bibitem[{\citenamefont{Chen and Xiao}(2008)}]{chen2008entropy}
\bibinfo{author}{\bibfnamefont{Y.}~\bibnamefont{Chen}} \bibnamefont{and}
  \bibinfo{author}{\bibfnamefont{Y.}~\bibnamefont{Xiao}},
  \bibinfo{journal}{Physics Letters B} \textbf{\bibinfo{volume}{662}},
  \bibinfo{pages}{71} (\bibinfo{year}{2008}).

\bibitem[{\citenamefont{Bekenstein}(2004)}]{bekenstein404042does}
\bibinfo{author}{\bibfnamefont{J.}~\bibnamefont{Bekenstein}},
  \bibinfo{journal}{Arxiv preprint quant-ph/0404042}  (\bibinfo{year}{2004}).

\bibitem[{\citenamefont{Schiffer and Bekenstein}(1989)}]{schiffer1989proof}
\bibinfo{author}{\bibfnamefont{M.}~\bibnamefont{Schiffer}} \bibnamefont{and}
  \bibinfo{author}{\bibfnamefont{J.}~\bibnamefont{Bekenstein}},
  \bibinfo{journal}{Physical Review D} \textbf{\bibinfo{volume}{39}},
  \bibinfo{pages}{1109} (\bibinfo{year}{1989}).

\bibitem[{\citenamefont{Padmanabhan}(2010)}]{padmanabhan2010thermodynamical}
\bibinfo{author}{\bibfnamefont{T.}~\bibnamefont{Padmanabhan}},
  \bibinfo{journal}{Reports on Progress in Physics}
  \textbf{\bibinfo{volume}{73}}, \bibinfo{pages}{046901}
  (\bibinfo{year}{2010}).

\bibitem[{\citenamefont{Marolf et~al.}(2004)\citenamefont{Marolf, Minic, and
  Ross}}]{marolf2004notes}
\bibinfo{author}{\bibfnamefont{D.}~\bibnamefont{Marolf}},
  \bibinfo{author}{\bibfnamefont{D.}~\bibnamefont{Minic}}, \bibnamefont{and}
  \bibinfo{author}{\bibfnamefont{S.}~\bibnamefont{Ross}},
  \bibinfo{journal}{Physical Review D} \textbf{\bibinfo{volume}{69}},
  \bibinfo{pages}{64006} (\bibinfo{year}{2004}).

\bibitem[{\citenamefont{Wald}(1999)}]{wald1999gravitation}
\bibinfo{author}{\bibfnamefont{R.}~\bibnamefont{Wald}},
  \bibinfo{journal}{Classical and Quantum Gravity}
  \textbf{\bibinfo{volume}{16}}, \bibinfo{pages}{A177} (\bibinfo{year}{1999}).

\bibitem[{\citenamefont{Abreu and Visser}(2010)}]{Abreu:2010sc}
\bibinfo{author}{\bibfnamefont{G.}~\bibnamefont{Abreu}} \bibnamefont{and}
  \bibinfo{author}{\bibfnamefont{M.}~\bibnamefont{Visser}}
  (\bibinfo{year}{2010}), \eprint{1005.1132}.

\end{thebibliography}
\end{acknowledgments}

\end{document}